\documentclass[prl,twocolumn,showpacs,superscriptaddress]{revtex4}
\usepackage{graphicx}
\begin{document}

\title{Enhanced Fusion-Evaporation Cross Sections in
Neutron-Rich $^{132}$Sn on $^{64}$Ni}
\date{\today}

\author{J.~F.~Liang}
\affiliation{Physics Division, Oak Ridge National Laboratory, Oak Ridge,
Tennessee 37831}
\author{D.~Shapira}
\affiliation{Physics Division, Oak Ridge National Laboratory, Oak Ridge,
Tennessee 37831}
\author{C.~J.~Gross}
\affiliation{Physics Division, Oak Ridge National Laboratory, Oak Ridge,
Tennessee 37831}
\author{J.~R.~Beene}
\affiliation{Physics Division, Oak Ridge National Laboratory, Oak Ridge,
Tennessee 37831}
\author{J.~D.~Bierman}
\affiliation{Physics Department AD-51, Gonzaga University, Spokane, Washington
99258-0051}
\author{A.~Galindo-Uribarri}
\affiliation{Physics Division, Oak Ridge National Laboratory, Oak Ridge,
Tennessee 37831}
\author{J.~Gomez~del~Campo}
\affiliation{Physics Division, Oak Ridge National Laboratory, Oak Ridge,
Tennessee 37831}
\author{P.~A.~Hausladen}
\affiliation{Physics Division, Oak Ridge National Laboratory, Oak Ridge,
Tennessee 37831}
\author{Y.~Larochelle}
\affiliation{Department of Physics and Astronomy, University of Tennessee,
Knoxville, Tennessee 37966}
\author{W.~Loveland}
\affiliation{Department of Chemistry, Oregon State University, Corvallis,
Oregon 97331}
\author{P.~E.~Mueller}
\affiliation{Physics Division, Oak Ridge National Laboratory, Oak Ridge,
Tennessee 37831}
\author{D.~Peterson}
\affiliation{Department of Chemistry, Oregon State University, Corvallis,
Oregon 97331}
\author{D.~C.~Radford}
\affiliation{Physics Division, Oak Ridge National Laboratory, Oak Ridge,
Tennessee 37831}
\author{D.~W.~Stracener}
\affiliation{Physics Division, Oak Ridge National Laboratory, Oak Ridge,
Tennessee 37831}
\author{R.~L.~Varner}
\affiliation{Physics Division, Oak Ridge National Laboratory, Oak Ridge,
Tennessee 37831}

\begin{abstract}
Evaporation residue cross sections have been measured with neutron-rich
radioactive $^{132}$Sn
beams on $^{64}$Ni in the vicinity of the Coulomb barrier. The average
beam intensity was  $2\times 10^{4}$ particles per second and the smallest
cross section measured was less than 5 mb. Large subbarrier fusion enhancement
was observed. Coupled-channels calculations taking into account inelastic
excitation and neutron transfer underpredict the measured cross sections 
below the barrier.
\end{abstract}


\maketitle

The interaction of two colliding nuclei consists of
an attractive nuclear potential and a repulsive Coulomb potential.
This creates a Coulomb barrier which the system
has to overcome in order to fuse. At energies below the barrier, fusion occurs
by quantum tunneling. Subbarrier fusion cross sections for
heavy ions are often found enhanced over the one-dimensional barrier
penetration model (BPM) prediction.
The enhancement can be explained in most cases by the coupling
of the relative motion and the nuclear structure degrees of freedom of the
participating nuclei\cite{be88}. It has been suggested that the fusion
yield would be further enhanced when the reaction is induced by unstable
neutron-rich nuclei\cite{ta92,hu91,da92}. This is attributed to the large N/Z
ratio of these nuclei reducing the barrier height and the
presence of a large number of nucleon transfer channels which can serve as
doorways to fusion\cite{ki94}. The
compound nucleus can be formed at lower excitation energies and with a smaller
fissility and, therefore, a higher survival probability for the
evaporation residues (ERs). Subbarrier fusion can be used
in experiments to produce superheavy elements.
It is beneficial to produce them near the
neutron shell closure to increase their lifetimes. If neutron-rich radioactive
beams are used in such experiments, heavy ions as projectiles
are required, since it is not possible to find a suitable target with light
ion beams.

The experimental search for fusion enhancement in heavy ion reactions has been
pursued at several laboratories using neutron-rich radioactive beams.
The measurements of $^{38}$S+$^{181}$Ta\cite{zy97} and
$^{29,31}$Al+$^{197}$Au\cite{si01} found only the enhancement expected
from the lowering of the barrier height caused by the larger radii of the
neutron-rich nuclei compared to
the stable $^{32}$S and $^{27}$Al, respectively.
This paper reports the first reaction study using accelerated unstable
neutron-rich $^{132}$Sn beams to measure fusion-evaporation cross sections.
The doubly magic $^{132}$Sn (Z=50, N=82) has eight extra neutrons compared
to the heaviest 
stable Sn isotope, $^{124}$Sn. The N/Z ratio of $^{132}$Sn (1.64) is larger
than that of $^{48}$Ca (1.4) and $^{208}$Pb (1.54) which are closed shell
nuclei commonly used to produce heavy elements\cite{ho00}.
The target, $^{64}$Ni, is semi-magic (Z=28) and
is the most neutron-rich stable isotope of nickel. The compound nucleus formed
in this experiment, $^{196}$Pt, lies in the valley of $\beta$-stability.
It has excitation energies greater than 30 MeV and can decay by particle
evaporation or fission.

The experiment was carried out at the Holifield Radioactive Ion Beam Facility
(HRIBF) at Oak Ridge National Laboratory. The isotope separator on-line
technique was used to produce radioactive $^{132}$Sn. A uranium
carbide target was bombarded by a 42 MeV proton beam 
accelerated by the Oak Ridge isochronous cyclotron.
Isobaric contaminants at A=132 were suppressed
by extracting molecular SnS$^{+}$ from the ion source and subsequently
breaking it up in the charge exchange cell where the SnS$^{+}$ was converted
to Sn$^{-}$\cite{st03} for injecting into the 25 MV tandem post accelerator.
The $^{132}$Sn ions were accelerated to six energies (453, 475, 489,
504, 536, and 560 MeV) and delivered to the target.
The beam intensity was measured by passing it
through a 10 $\mu$g/cm$^{2}$ carbon foil and detecting the secondary electrons
in a microchannel plate (MCP) detector. Three of these MCP systems were used
in this experiment for monitoring the beam and providing timing signals. The
average beam intensity was 2$\times 10^{4}$ particles per second (pps) with
a maximum near 3$\times 10^{4}$ pps. The
purity of the $^{132}$Sn beam was checked by measuring the energy
loss in an ionization chamber (IC). A $^{132}$Te beam was used to calibrate
the energy loss spectrum. It was determined that the impurity was less than
2\% and that all measurable impurities had a higher atomic number (Z) than Sn.
This impurity has negligible effect on the
measurement because the higher Coulomb barrier suppresses the fusion of the
contaminants with the target. A $^{124}$Sn beam was used as a guide beam
to set up the accelerator and beam line optics. At the target position, the
beam was focused to a spot 1.0 mm horizontally and 2.5 mm vertically. The shape
of the guide beam was recorded by an electronic phosphor\cite{sh98} consisting
of an aluminized mylar foil and a position sensitive microchannel plate
detector located 74 cm in front of the target. This beam was also used for
testing the detector system.
The $^{132}$Sn beam was then tuned by scaling the optical elements
and comparing the beam shape with that of the guide beam using the
electronic phosphor.

\begin{figure}[h]
\includegraphics[width=2.25in]{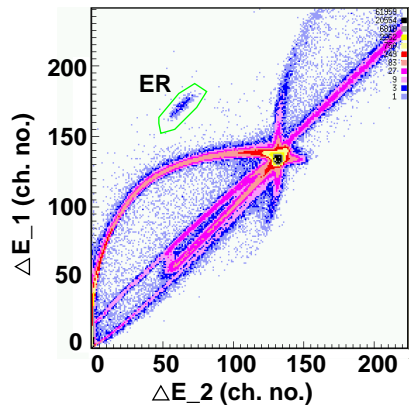}
\caption{\label{fg:ede}Histogram of the energy loss of beam and ERs measured
in the first two anodes of the ionization chamber for 536 MeV
$^{132}$Sn+$^{64}$Ni.}
\end{figure}

The ERs were detected along with beam particles by a timing detector and an
IC located 16.9 cm from the target at 0$^{\circ}$. They
were identified by their time-of-flight and energy
loss in the IC. The acceptance of the
timing detector was a 2.54 cm diameter circle and the detection efficiency
was approximately 100\% for these heavy ions. In the time-of-flight
measurement, the coincidence between two timing detectors placed 119 and
315 cm upstream from the target provided the timing references. The data
acquisition was triggered by the scaled down beam singles or the ER-beam
particle coincidences. With this triggering scheme an overall deadtime of
less than 5\% was achieved. The IC was filled with
CF$_{4}$ gas. The pressure was adjusted between 50 and 60 Torr to optimize the
separation of ERs from the beam. 
A detailed description of the experimental
apparatus will be published elsewhere\cite{sh03}. Figure~\ref{fg:ede} shows
the histogram of the energy loss in the first two segments of the IC
for a beam energy of 536 MeV. Although there is some signal pile-up
introduced by directly injecting the beam into the detector, it is clear that
the ERs are still well separated from the beam. With this setup, measurement
of ER cross sections less than 5 mb can be achieved.

The cross section was obtained by integrating the ER yield and
summing the beam particles in the IC.
Because of the low intensity of radioactive beams, the measurement was
performed with a thick target, 1 mg/cm$^{2}$ self-supporting highly enriched
(99.81\%) $^{64}$Ni foil.
The target thickness was determined by measuring the energy loss of $\alpha$
particles emitted from a $^{244}$Cm source and a 536 MeV $^{132}$Sn beam
passing through the target, and by measuring the weight and area of the target.
The energy loss of $^{132}$Sn projectiles in the target was approximately
40~MeV. At energies below the
Coulomb barrier, the excitation function falls off exponentially. For this
reason, the measured cross section ($\sigma_{meas}$) is sensitive
predominantly to the front portion of the target and is actually a
weighted average of the cross section over the range of energy loss in the
target, from the energy
of the beam entering the target ($E_{in}$) to that exiting the target 
($E_{out}$), namely,
\begin{equation}
\sigma_{meas} = \frac{\int_{E_{in}}^{E_{out}} \sigma (E)dE}
{\int_{E_{in}}^{E_{out}} dE}.       \label{eq:avsig}
\end{equation}
To determine the effective reaction energy, the cross section
was parameterized as an exponential function,
$\sigma(E) = N\exp (\alpha E)$,
where $N$ is a normalization factor and $\alpha$ is a slope parameter. 
By solving the integral equation~(\ref{eq:avsig}) for two adjacent data
points in the excitation function, $N$ and $\alpha$ were obtained.
Subsequently, the effective energy is deduced by inverting the exponential
function, namely, $E = ln(\sigma_{meas}/N)/\alpha$ .

Since this experiment was performed in inverse kinematics (a heavy projectile
on a light target) the ERs were very forward focused. However, the shape of
the beam spot was not symmetric. Moreover, one of the disadvantages of using a 
thick target is the multiple scattering which
results in broadening the angular distribution. Monte Carlo simulations were
used to estimate the efficiency of the apparatus. The angular distribution of
ERs was generated by the statistical model code {\sc pace}\cite{ga80} and
the width of the distribution of
multiple scattering angles was predicted by Ref~\cite{ma67}. The simulations
show that the efficiency of the apparatus changes from $95\pm 1$\% for
the lowest beam energy to $98\pm 1$\% for the highest energy.

\begin{figure}[h]
\includegraphics[width=2.25in]{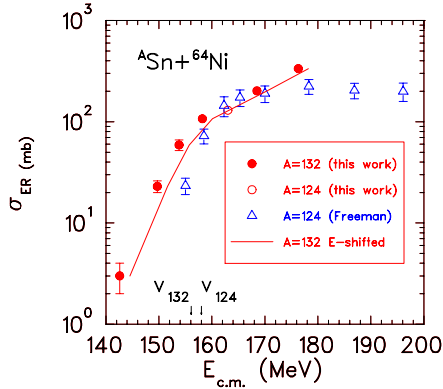}
\caption{\label{fg:132-124}Fusion-evaporation excitation functions of
$^{132}$Sn+$^{64}$Ni (filled circles) and $^{64}$Ni+$^{124}$Sn\cite{fr83}
(open triangles). The open circle is our measurement using a $^{124}$Sn beam.
The solid curve is the excitation function of
$^{132}$Sn+$^{64}$Ni shifted in energy by the ratio of nuclear radii relative
to $^{64}$Ni+$^{124}$Sn.}
\end{figure}

Figure~\ref{fg:132-124} presents the fusion-evaporation excitation function
of $^{132}$Sn+$^{64}$Ni measured in this work (solid circles) and that of
$^{64}$Ni+$^{124}$Sn measured by Freeman {\em et al.}\cite{fr83}
(open triangle). The open circle is our measurement using the $^{124}$Sn guide
beam which is consistent with the measurement in Ref.~\cite{fr83}. The
uncertainty in energy for our data is smaller than the size of the
symbol in the figure. To compare
the two excitation functions, the energy of $^{132}$Sn+$^{64}$Ni is shifted
by the difference between the fusion barriers for the two systems predicted by
the Bass model\cite{ba74} ($\Delta V_{B} = 1.93$~MeV).
The energy shifted excitation function of $^{132}$Sn+$^{64}$Ni is
shown by the solid curve in Fig.~\ref{fg:132-124}. It can be seen that at
the highest energy the ER cross sections for
$^{132}$Sn+$^{64}$Ni are larger. This can be expected from the higher
stability against fission for the neutron-rich compound nucleus.
At energies below the barrier, the ER cross sections
for $^{132}$Sn+$^{64}$Ni are found much enhanced comparing to those of
$^{64}$Ni+$^{124}$Sn and a simple shift of the barrier height cannot explain
the enhancement. 

To compare the measured excitation function with fusion models, it is necessary
to estimate fission yields in the reaction. Statistical model calculations
were carried out using the code {\sc pace}. The input parameters were
determined by reproducing the ER and fission cross sections of
$^{64}$Ni+$^{124}$Sn in Ref.~\cite{le86}. The following parameters were
used: level density parameter $a=A/8$~MeV$^{-1}$
where $A$ is the mass of the compound nucleus, ratio of the Fermi gas level
density parameter at the saddle point to that of the ground state 
$a_{f}/a_{n}=1$, diffuseness of spin distribution $d=4 \hbar$, and Sierk's
fission barrier\cite{si86}. 
The calculations predict that fission is negligible for
$^{132}$Sn+$^{64}$Ni and $^{64}$Ni+$^{124}$Sn at E$_{\rm c.m.} \leq 160$~MeV.
Therefore, the following discussion will be restricted to the data points at
E$_{\rm c.m.} \leq 160$~MeV where the ER cross sections are taken as fusion
cross sections.

\begin{figure}[h]
\includegraphics[width=2.25in]{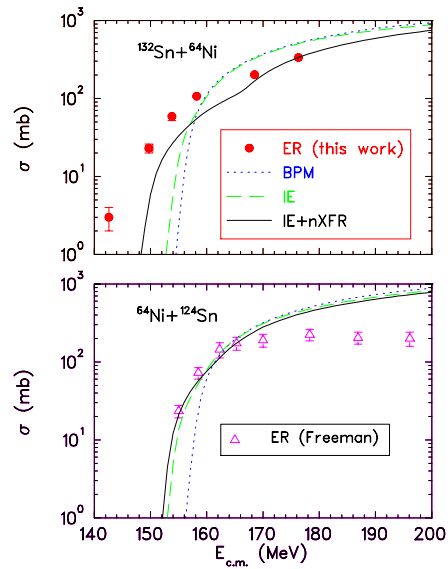}
\caption{\label{fg:ccfull}Comparison of measured ER excitation functions with
fusion model calculations. The upper panel is for $^{132}$Sn+$^{64}$Ni and the
lower panel is for $^{64}$Ni+$^{124}$Sn\cite{fr83}. The measured ER cross
sections are shown by the filled circles and open triangles for
$^{132}$Sn+$^{64}$Ni and $^{64}$Ni+$^{124}$Sn, respectively. The one
dimensional barrier penetration model (BPM) calculation is shown by the
dotted curve. The
dashed and solid curve are results of coupled-channels calculations including
inelastic excitation (IE), and IE and neutron transfer (nXFR), respectively.}
\end{figure}

Large subbarrier fusion enhancement in $^{132}$Sn+$^{64}$Ni can be seen when
the excitation function
is compared to a one-dimensional BPM shown by the
dotted curve in the upper panel of Fig.~\ref{fg:ccfull}. The nuclear potential
was assumed to have a Woods-Saxon shape. The potential parameters were
obtained by adjusting them to reproduce the fusion cross section
of $^{64}$Ni+$^{124}$Sn in Ref.~\cite{le86} at high energies. 
They are: depth $V_{\circ} = 76.6$ MeV, radius
parameter $r_{\circ} = 1.2$ fm, and diffuseness parameter $a = 0.65$ fm.

It is well established that
subbarrier fusion enhancement can be described by channel
couplings\cite{be88}. The couplings
result in splitting the single barrier into a distribution of barriers.
The incident flux overcoming the low energy barriers gives rise to the enhanced
fusion cross sections\cite{ro91,da98,ba98}.
Coupled-channels calculations were performed with the code
{\sc ccfull}\cite{ha99} which takes into account the effects of nonlinear
coupling to all orders. The calculations used the same nuclear potential
as that for the BPM calculation. The dashed curves in
Fig.~\ref{fg:ccfull} is the result of coupling to inelastic excitation (IE)
of the projectile and target. Table~\ref{tb:cc} lists the states and
parameters\cite{ra87,sp89} for the calculations. As shown in the lower panel
of Fig.~\ref{fg:ccfull}, the calculation reproduces the
$^{64}$Ni+$^{124}$Sn cross sections fairly well at low energies.
For $^{132}$Sn+$^{64}$Ni, the calculation significantly underpredicts the
subbarrier cross sections as shown in the upper panel of Fig.~\ref{fg:ccfull}. 
The small effect of coupling to IE
in $^{132}$Sn can be attributed to the high excitation
energy of the 2$^{+}$ excited state and the small reduced transition
probability (B(E2)).
\begin{table}
\caption{\label{tb:cc}Parameters used in coupled-channels calculations.
$\lambda^{\pi}$ is the angular momentum and parity and $\beta_{\lambda}$ is the
deformation parameter.}
\begin{ruledtabular}
\begin{tabular}{cccc}
Nucleus & $\lambda^{\pi}$       &E$^{*}$ & $\beta_{\lambda}$    \\ \hline
$^{64}$Ni       & 2$^{+}$       & 1.346 & 0.179         \\
$^{124}$Sn      & 2$^{+}$       & 1.132 & 0.095 \\
                & 3$^{-}$       & 2.614 & 0.136 \\
$^{132}$Sn      & 2$^{+}$       & 4.041 & 0.06  \\
\end{tabular}
\end{ruledtabular}
\end{table}

In $^{64}$Ni+$^{124}$Sn,
the two neutron transfer reaction is the only transfer channel which has a
positive Q-value. Coupled-channels calculations including this channel with
an empirically determined coupling constant of 0.25 MeV and IE
are in good agreement with the fusion cross
sections near and below the barrier, as can be seen by the solid curve in the
lower panel of Fig.~\ref{fg:ccfull}. It is 
noted that the code {\sc ccfull} is suitable for reactions where
multinucleon transfer is less important than IE\cite{ha99}
as is the case in $^{64}$Ni+$^{124}$Sn.
For the $^{132}$Sn-induced reaction, the Q-values are positive for transfer of
two to six neutrons. This large number of neutron transfer channels with
positive Q-values and the small effects of IE channel
coupling suggest that multi-neutron transfer could play an important role in
the observed fusion enhancement. Although {\sc ccfull} is not expected to
treat the coupling of multinucleon transfer accurately, exploratory
calculations were carried out to provide preliminary estimate of the effects
of coupling to these
channels. Results of calculations including these transfer channels and IE
are shown by the solid curve in the upper panel of Fig.~\ref{fg:ccfull}. The
calculation cannot account for the cross sections near and below the barrier,
nevertheless, it illustrates qualitatively the enhancement of subbarrier
fusion due to the coupling to multinucleon transfer.
It would be interesting to study near-barrier fusion further using even more
neutron-rich Sn isotopes. However, this will be a very challenging
task because the present beam intensity for $^{134}$Sn at HRIBF is
approximately 2000 pps and highly contaminated. On the other hand, HRIBF can
provide other pure neutron-rich radioactive beams such as Br and I with
reasonable intensities for further studies.

In the future, it is necessary to measure fission for $^{132}$Sn+$^{64}$Ni
in order to obtain the fusion cross sections and study the survival probability
of the compound nucleus. In addition, it was found that
the extra-push energy\cite{sw81} is needed for compound nucleus formation
in $^{64}$Ni on stable even Sn isotopes at high energies\cite{le86} and
the extra-push energy diminishes as the number of neutrons in Sn increases.
The threshold for requiring the extra-push energy given in
Ref.~\cite{le86} is near the $^{132}$Sn+$^{64}$Ni system. This can be
investigated by measuring ER and fission cross sections at higher energies.

In summary, fusion-evaporation cross sections using neutron-rich 
$^{132}$Sn beams on a $^{64}$Ni target were measured at energies near the
Coulomb
barrier. At the highest energy the ER cross section is larger
than that of $^{64}$Ni on stable $^{124}$Sn. This is expected since the extra
neutrons increase the survival probability of the compound nucleus. However,
measurements of both ER and fission cross sections at higher energies will
illustrate this better. Large subbarrier fusion enhancement
using neutron-rich radioactive heavy ion
beams was observed in this experiment. The subbarrier enhancement cannot be
explained by a simple shift of the barrier height, or by channel couplings
to inelastic excitation channels.
There are five neutron transfer channels which
have large positive Q-values. These reaction channels may serve as doorways
to fusion.
Further experiments using neutron-rich radioactive beams would advance our
understanding of the
mechanism for the fusion enhancement, and provide valuable information
for using such beams to produce superheavy elements at future radioactive
beam facilities.

We wish to thank the HRIBF staff for providing excellent radioactive beams and
technical support. We wish to thank S. Novotny for his help with the
apparatus. Research at the Oak Ridge National Laboratory is 
supported by the U.S. Department of Energy under contract DE-AC05-00OR22725 
with UT-Battelle, LLC. W.L. and D.P. are supported by the the U.S. Department
of Energy under grant no. DE-FG06-97ER41026.

\end{document}